\documentclass[aps,prl,superscriptaddress,twocolumn]{revtex4-2}
\usepackage{amsmath}
\usepackage{graphicx}	
\usepackage{bm} 
\usepackage{color}
\usepackage{setspace}
\usepackage{placeins}
\usepackage[dvipsnames]{xcolor}
\usepackage{mathtools}
\usepackage{pdfpages}
\usepackage{etoolbox} 

\makeatletter
\patchcmd{\@outputpage@head}{\@ifx{\LS@rot\@undefined}{}{\LS@rot}}{}{}{}
\makeatother

\begin{document}
\title{Connecting scrambling and work statistics for
short-range interactions in the harmonic oscillator}
\date{\today}
\author{M.~Mikkelsen}
\email[]{mathias-mikkelsen@phys.kindai.ac.jp}
\affiliation{Quantum Systems Unit, OIST Graduate University, Onna, Okinawa 904-0495, Japan}
\affiliation{Department of Physics, Kindai University, Higashi-Osaka City, Osaka 577-8502, Japan}
\author{T.~Fogarty}
\affiliation{Quantum Systems Unit, OIST Graduate University, Onna, Okinawa 904-0495, Japan}
\author{Th.~Busch}
\affiliation{Quantum Systems Unit, OIST Graduate University, Onna, Okinawa 904-0495, Japan}
 
\begin{abstract} 
We investigate the relationship between information scrambling and work statistics after a quench for the paradigmatic example of short-range interacting particles in a one-dimensional harmonic trap, considering  up to five particles numerically. In particular, we find that scrambling requires finite interactions, in the presence of which the long-time average of the squared commutator for the individual canonical operators is directly proportional to the variance of the work probability distribution. In addition to the numerical results, we outline the mathematical structure of the $N$-body system which leads to this outcome. We thereby establish a connection between the scrambling properties and the induced work fluctuations, with the latter being an experimental observable that is directly accessible in modern cold atom experiments.
\end{abstract}

\maketitle
The response to a sudden change in the Hamiltonian is a topic which has led to many valuable insights into the physics of quantum many-particle systems in recent years. Quenches have been used to probe phase transitions \cite{Silva2008,Karrasch2013,Campbell2016,Heyl2017,Fogarty2017,Sengstock2017,Mikkelsen2018}, explore the orthogonality catastrophe \cite{Campo2011,Goold2011,Cetina2015, Parish2016, Demler2016, Demler2018} and investigate irreversibility, thermodynamics and equilibration properties \cite{Rigol2007,Rigol2008,Rigol2016,Vidmar2016,March2016}. For example systems that obey the eigenstate thermalization hypothesis have been shown to thermalize \cite{Rigol2008,Rigol2016} while integrable systems do not \cite{Rigol2007,Vidmar2016}. Studying the dynamical response of a system to a sudden quench usually entails calculating the time-evolution of expectation values of observables such as the momentum distribution. However, one can also characterize a quench more broadly, for example through operator-independent (aside from the final Hamiltonian $\hat{H}_F$) quantities such as the diagonal ensemble \cite{Rigol2008,Polkovnikov2011} and the closely related experimentally measurable work probability distribution \cite{Campisi2011,Fusco2014,Serra2014,Cerisola2017,Keller2016,March2016}. The statistical moments of the work probability distribution, $\langle W^{\alpha} \rangle= \text{Tr} [(\hat{H}_F-\hat{H}_I)^{\alpha} \hat{\rho}_I]$, where $\hat{H}_I$ is the initial Hamiltonian and $\hat{\rho}_I$ is the initial state, are often used to give an indication of the irreversibility of the quench process \cite{Fusco2014}. One example of this is the irreversible work $\langle W_\text{irr}\rangle = \langle W \rangle-\Delta F$ which quantifies the disparity between the average work and the free energy during a non-quasi-static process. Further insight can be gained through the variance of nonequilibrium fluctuations about the average,  $ \Delta W^2 = \langle W^2\rangle  -\langle W \rangle^2 $, which is of interest in the field of statistical quantum thermodynamics \cite{Esposito2009,Campisi2011,Rigol2016} and has been suggested as a probe of critical behaviour \cite{Jaramillo2017,Nigro2019}. 

Since the work probability distribution is related to the delocalisation of the initial state in the Hilbert space defined by the eigenstates of the final Hamiltonian, it is natural to characterize this further by investigating the delocalisation dynamics. This process is often referred to as scrambling \cite{Swingle2018,Rey2019}, whereby over time the initial state can no longer be reconstructed from local measurements alone. One particular measure of this scrambling is the expectation value of the squared commutator of two operators $\hat{A}(t)=e^{i\hat{H}t}  \hat{A} e^{-i\hat{H}t}$ and $\hat{B}$, $C_{AB}(t)=\langle [\hat{A}(t),\hat{B} ]^2 \rangle$ \cite{Swingle2018}, which can be rewritten in terms of time-dependent correlation functions as
$C_{AB}(t) = D_{AB}(t)+I_{AB}(t)-2 \;\text{Re}[F_{AB}(t)]$, with
\begin{align}
D_{AB}(t) &= \langle \hat{B}^\dagger \hat{A}^\dagger(t) \hat{A}(t) \hat{B} \rangle\,, \label{eq:Dgeneric} \\
I_{AB}(t) &= \langle \hat{A}^\dagger(t) \hat{B}^\dagger \hat{B} \hat{A}(t) \rangle\,, \label{eq:Igeneric} \\
F_{AB}(t) &= \langle \hat{A}^\dagger(t) \hat{B}^\dagger  \hat{A}(t) \hat{B} \rangle\,. \label{eq:Fgeneric}
\end{align}
Most work in recent years has focused on the 4-point out-of-time ordered correlation function (4-OTOC) $F_{AB}(t)$, as $D_{AB}(t)$ is time-ordered and  $I_{AB}(t) = \langle \hat{A}^\dagger \hat{B}^\dagger(-t) \hat{B}(-t) \hat{A} \rangle$ is anti-time-ordered for an eigenstate of the Hamiltonian. The squared commutator and the 4-OTOC were initially proposed as measures of quantum chaos \cite{Stanford2016} but have recently been shown to be powerful tools for studying information scrambling in non-chaotic systems as well, for example near quantum critical points \cite{Heyl2018,Lin2018,Dag2019}, in the presence of many-body entanglement and coherence \cite{Rey2017,McGineley2019}, and in quantum thermodynamics \cite{Goold2017,Swan2019}. For initial states that are not eigenstates, e.g.~states after a quench, $I_{AB}(t)$ is also not time-ordered and called a 3-point OTOC (3-OTOC) \cite{Hamazaki2018}. One can see that $I_{AB}(t)$ is readily interpretable as a time-reversal test, i.e.~it corresponds to taking the expectation value of $\hat{B}^\dagger \hat{B}$ with the quantum-state $\hat{A}(t)|\psi \rangle$. It therefore measures how much the time-reversal symmetry is broken by the application of the operator $\hat{A}$. 

While in discrete systems schemes for measuring the OTOCs have been experimentally implemented using a time-reversal protocol \cite{Rey2017}, in continuum systems such a direct implementation is extremely difficult as it requires reversing the kinetic energy terms. Therefore, finding a connection between information scrambling and other measures of irreversibility, particularly ones that can be measured in continuum systems is important. While progress towards such an understanding has recently been made
\cite{Halpern2017,Goold2017,Hamazaki2018,Campo2019,Yan2020}, we will focus in this work on non-chaotic systems and look at experimentally available cold-atom systems of interacting bosons in quasi-one-dimensional traps. Such systems offer an ideal testbed to study non-equilibrium dynamics as advances in the experimental manipulation of single- and few-body systems allows for precise control over their interactions and trapping potentials \cite{Jochim2012,Endres2016}. The total number of particles can also be tuned deterministically allowing one to explore the cross-over between few- and many-body physics \cite{Wenz2013}. They are therefore highly suitable to consider how information scrambling emerges after sudden quenches, specifically as a function of finite interactions between the particles.

The system we consider consists of $N$ particles and can be described by the  dimensionless Hamiltonian
\begin{align}
\hat{H} = \sum_{j=1}^N \left[ -\frac{1}{2}\frac{\partial^2}{\partial \hat{x}^2_j}+\frac{1}{2}  \Omega^2(t) \hat{x}_j^2 \right] + \sum_{k>j} g \, \delta (\hat{x}_k-\hat{x}_j)\;,
\label{eq:Hharmonictrap}
\end{align}
where the interactions and trap frequency are parametrized by $g$ and $\Omega(t)$, respectively. To explore nonequilibrium scrambling in this system we consider the canonical operators, $\hat{x}_j$ and $\hat{p}_j$, after a sudden change of the trapping potential described by $\Omega(t)=\gamma+\Theta(t)(1-\gamma)$, where $\Theta(t)$ is the Heaviside step function. The trap strength in the initial Hamiltonian is therefore given by $\gamma$, while the final Hamiltonian has a trap strength of unity. This allows us to scale all relevant quantities in units of the final Hamiltonian and all results only depend on $\gamma$, which then quantifies the strength of the quench and whether the trap is compressed ($\gamma<1$) or expanded ($\gamma>1$). We keep the interaction strength fixed throughout the dynamics with $g>0$ describing repulsive interactions. This allows us to clearly identify the effects of finite interactions on the information scrambling and work statistics after the quench of the trapping potential.

Sudden quenches are characterized by the eigenspace of the final Hamiltonian $\hat{H}_F | \psi_j \rangle = \ E_j | \psi_j \rangle$ and the overlap coefficients $c_{j} = \langle \psi_j | \psi^I \rangle$, where $| \psi^I \rangle$ is the initial state with energy $E^I$. This allows one to write the contributions to the squared commutator as
\begin{eqnarray}
D_{AB}(t) &=& \sum_{\mathclap{j,k,n,m}} c_{j}^* c_{k} e^{-i(E_{mn})t}B^\dagger_{jn}\langle \hat{A}^\dagger \hat{A} \rangle_{nm} B_{mk}   \label{eq:Dgenericexplicit} ,\\
I_{AB}(t) &=& \sum_{\mathclap{j,k,n,m}} c_{j}^* c_{k} e^{-i(E_{kj}+E_{nm})t}A^\dagger_{jn}\langle \hat{B}^\dagger \hat{B} \rangle_{nm} A_{mk}   \label{eq:Igenericexplicit} ,\\
F_{AB}(t) &=& \sum_{\mathclap{j,k,n,m}} c_{j}^* b_{k} e^{-i(E_{kj}+E_{nm})t}A^\dagger_{jn} B^\dagger_{nm} A_{mk} ,  \label{eq:Fgenericexplicit}
\end{eqnarray}
where  $b_{j} = \langle \psi_j | \hat{B}| \psi^I \rangle$, $A_{jk}=\langle \psi_j | \hat{A} | \psi_k \rangle$, $\langle \hat{A}^\dagger \hat{A} \rangle_{nm}=\langle \psi_n | \hat{A}^\dagger \hat{A} | \psi_m \rangle$, and the other operator matrix elements are defined similarly. The energy differences are given by $E_{mn}=E_m-E_n$. The statistical moments of the work probability distribution can be expressed as $\langle W^{\alpha} \rangle=\sum_j |c_j|^{2} (E_j-E^I)^{\alpha}$ with $\alpha=1,2,\dots$ \cite{Fusco2014}. The variance $ \Delta W^2 = \langle W^2\rangle  -\langle W \rangle^2 $ will be used as a quantifier of the irreversibility of the quench dynamics, while the information scrambling will be gauged by the infinite time-average of the squared commutator $\bar{C}_{AB}=\lim_{T \rightarrow \infty}\frac{1}{T}\int_{0}^T \langle [\hat{A}(t),\hat{B} ]^2 \rangle dt$. Time-averaged behaviour has recently attracted more attention and has been connected to the description of quantum phases \cite{Heyl2018,Dag2019}. 

The Hamiltonian in Eq.~\eqref{eq:Hharmonictrap} possesses analytical many-body solutions in the non-interacting limit $g=0$ and the Tonks-Girardeau (TG) limit of infinite repulsive interactions, $g\rightarrow \infty$. In both cases the many-body system is described by a harmonic spectrum which elicits self-similar dynamics after changes to the trapping frequency \cite{Minguzzi2005,Atas2017}, and the scrambling of canonical operators in these limits therefore simply reflects the single-particle breathing mode following a trap quench. In fact, it can be shown that the time averaged scrambling in both limits for the individual canonical operators $[\hat{A}_i(t),\hat{B}_j]^2$, where  $\hat{A}_i=\hat{x}_i,\hat{p}_i$ and $\hat{B}_j=\hat{x}_j,\hat{p}_j$ is given by \cite{SuppMat}  $\bar{C}_{A_j, B_k}^{g=0}=\bar{C}_{A_j, B_k}^{g\rightarrow\infty}=\frac{1}{2} \delta_{jk}$. The scrambling is therefore independent of both the system-size and the strength of the trap quench $\gamma$. 

For finite interactions, $g>0$, the energy levels acquire non-trivial shifts $E^{g>0}_j=E^{g=0}_j+\Delta_j$ which destroy the regularity of the harmonic oscillator spectrum. This leads to complex dynamics which do not admit a single particle description and introduces correlations between the particles. While large systems become computationally intractable, few-body systems are solvable while retaining the physics stemming from the finite contact interactions \cite{Sowinski2019}. For $N=2$ particles analytic solutions exist \cite{Busch98,Schmelcher2019} which can be used to find an analytic expression for the full squared commutator \cite{SuppMat}. For larger systems, $N=3,4,5$, one must solve the Hamiltonian in Eq.~\eqref{eq:Hharmonictrap} numerically which we do by utilizing exact diagonalization techniques \cite{March2013} with an effective interaction approach \cite{Lindgren2014} and an optimized choice of the many-body basis \cite{Plodzien2018}. 
We will focus on the dynamics of $[\hat{x}_1(t),\hat{x}_1]^2$, as other combinations of canonical operators give similar results \cite{SuppMat}. 

\begin{figure}
\centering
\includegraphics[width=1\linewidth]{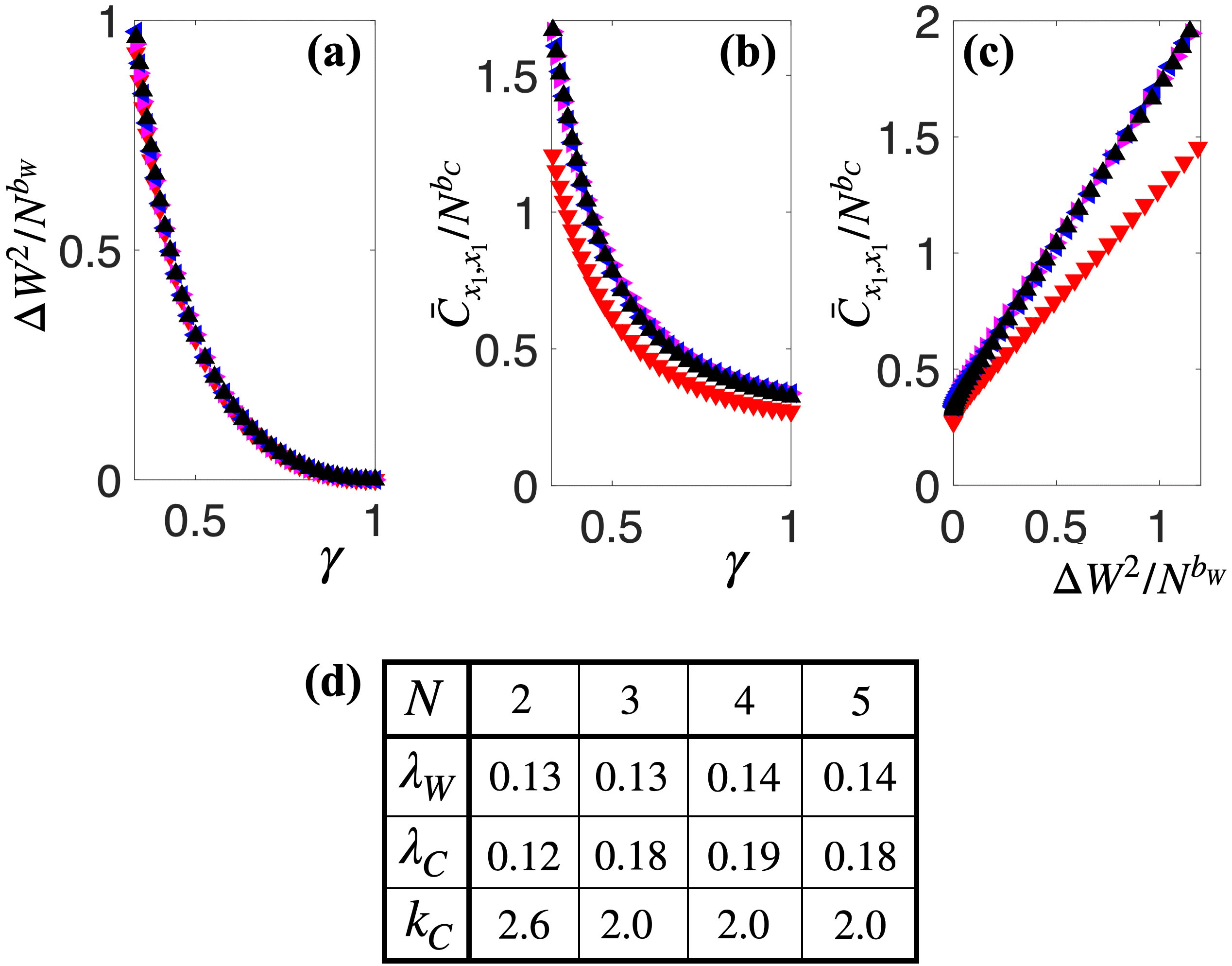}
\caption{(a) Variance and (b) time-averaged squared commutator as a function of the quench strength $\gamma$ for $g=5$. (c) Time-averaged squared commutator as a function of the variance. The colors correspond to $N=2$ (red), $N=3$ (magenta), $N=4$ (blue) and $N=5$ (black). The time-averaged squared commutator is scaled with $N^{b_C}$, while the the variance is scaled with $N^{b_W}$, with  $b_C(5) =1.7$ and $b_W(5) =2$. (d) Corresponding fitting components as a function of particle number.}
\label{fig:Systemsizecomparison} 
\end{figure}

In Fig.~\ref{fig:Systemsizecomparison}(a,b) the variance of the work distribution and average scrambling is shown as a function of the quench strength $\gamma$ for finite interactions $g=5$ and different system sizes. Reducing $\gamma$ (increasing the compression of the trap) drives the system further from equilibrium and therefore increases both the variance and the information scrambling. For the different system sizes the variance and the average scrambling are rescaled by $N^{b_W}$ and $N^{b_C}$, respectively, where the exponents $b_W$ and $b_C$ are found by extrapolating the behaviour of the system in the analytically solvable limits $g=\{0,\infty\}$. In these the variance as a function of $N$ and $\gamma$ is $\Delta W_{g=0}^2=\frac{N}{8}(\gamma-\frac{1}{\gamma})^2$ and $\Delta W_{g=\infty}^2=\frac{N(N^2+2)}{24}(\gamma-\frac{1}{\gamma})^2$, which evinces that the interaction only affects how the system size scales. For finite interactions we therefore fit the function $\Delta W_{g,N}^2=N^{b_W(g)}\lambda_W(g,N)(\gamma-\frac{1}{\gamma})^2$ with the exponent having values $1<b_W(g)<3$ which are $g$ dependent. Similarly for the time-averaged squared commutator the following function gives a good fit $\bar{C}_{x_1,x_1}=N^{b_C(g)}\lambda_C(g,N)[(\gamma-\frac{1}{\gamma})^2+k_C(g,N)]$. For up to $N=5$ particles the leading exponents of the system size are found to be $b_W(5)\approx 2$ and $b_C(5)\approx 1.7$.

In Fig.~\ref{fig:Systemsizecomparison}(d) we list the numerically obtained values of the remaining fitting constants showing that they quickly converge for $N\geq3$, which can also be seen in  Fig.\ref{fig:Systemsizecomparison}(a,b) as the data for $N=3,4,5$ show strong convergence. In Fig.\ref{fig:Systemsizecomparison}(c) we plot $\bar{C}_{x_1,x_1}$ as a function of the variance $\Delta W^2$, showing that the average information scrambling is linearly proportional to the work fluctuations. For a system with finite interactions the information scrambling is therefore closely related to the irreversible non-equilibrium excitations created by the trap quench, something which is absent in the $g=\{0,\infty\}$ limits where $\bar{C}_{x_1,x_1}=1/2$ and therefore does not depend on the system size $N$ or quench strength $\gamma$.

The results in Fig.~\ref{fig:Systemsizecomparison} can be explained in more detail by considering the structure of the Hamiltonian and the squared commutator analytically. To do this, we first outline some generic conditions under which the squared commutator will simplify and which are applicable to other similar models. Firstly, we consider a non-degenerate system [condition (\textit{i})], which is the case in many situations of interest. From Eqs.~(\ref{eq:Dgenericexplicit}-\ref{eq:Fgenericexplicit}) one can see that contributions to the infinite-time average of the squared commutator are only obtained when the complex exponential equals $1$, which means that $D_{AB}(t)$ has contributions whenever $E_m=E_n$. The contributions to $I_{AB}$ and $F_{AB}$ can be split into 3 cases: the energy-differences can be pairwise zero in the case where  $E_{k}=E_{j}$ and $E_m=E_n$ or the sum can be zero when $E_{k}=E_{m}$ and $E_{n}=E_{j}$. Finally, it is also possible that $E_k-E_j+E_n-E_m=0$ for $j\neq k \neq n \neq m$. However, we will consider an additional constraint on the spectrum, namely  $E_k-E_j+E_n-E_m \neq 0$ for $j\neq k \neq n \neq m$ [condition (\textit{ii})] which ensures that these terms have no contributions. This often holds in chaotic systems \cite{Srednicki1999}, although it needs to be explicitly shown for any system of interest. The resulting time-averages can therefore be written as
\begin{align}
\bar{D}_{AB} =& \sum_{j,k,n} c_{j}^* c_{k} B^\dagger_{jn}\langle \hat{A}^\dagger \hat{A} \rangle_{nn} B_{nk} , \label{eq:timeaverageDfull} \\
\bar{I}_{AB} =& \sum_{j,n} |c_{j}|^2 A^\dagger_{jn}\langle \hat{B}^\dagger \hat{B} \rangle_{nn} A_{nj}  \nonumber \\
&+\sum_{j\neq k} c_{j}^* c_{k} A^\dagger_{jj}\langle \hat{B}^\dagger \hat{B} \rangle_{jk} A_{kk}  \label{eq:timeaverageIfull} \\
\bar{F}_{AB} =& \sum_{j,n} c_{j}^* b_{j} A^\dagger_{jn} B^\dagger_{nn} A_{nj} \nonumber \\
&+\sum_{j\neq k} c_{j}^* b_{k} A^\dagger_{jj}B^\dagger_{jk} A_{kk}  \label{eq:timeaverageFfull}.
\end{align}

If the system obeys a final constraint on the matrix-elements of the operators with respect to the eigenbasis of the final Hamiltonian, namely $B_{kk}=A_{kk}=0$ [condition (\textit{iii})] the second term in Eq.~\ref{eq:timeaverageIfull} and both terms in Eq.~\ref{eq:timeaverageFfull} will be zero. This constraint is much less generic than those on the spectrum and is only fulfilled by certain classes of models and operators. In general it will be  obeyed by systems with an odd-even parity symmetry and for operators which change the parity of a state. 

The time-averages for any system which fulfills conditions (\textit{i})-(\textit{iii}) reduce to $\bar{F}_{AB}=0$ and $\bar{C}_{AB} = \bar{D}_{AB} + \bar{I}_{AB}$, with
\begin{align}
\bar{D}_{AB} = \sum_{j,k}c_{j}c_{k} K^{BA}_{jk} \; , \quad \quad \bar{I}_{AB} = \sum_{j}|c_{j}|^2 K_{jj}^{AB}\;,
\label{eq:timeaverageofIandD}
\end{align}  
where $K_{jk}^{AB}  = \sum_{n} A^\dagger_{jn}\langle \hat{B}^\dagger \hat{B} \rangle_{nn} A_{nk}$.
$\bar{I}_{AB}$ is given as the diagonal ensemble expectation value of an emergent operator and is therefore directly related to the work statistics of the quench with no dependence on the sign of the overlap coefficients. $\bar{D}_{AB}$ is given as a sum over all the off-diagonal values of a similar emergent operator which means that the sign of the overlap coefficients matter and negative and positive contributions can interfere destructively. 

The symmetries of a system of $N$ interacting particles in a harmonic trap have been thoroughly explored \cite{Sowinski2019,Harshman2012,Harshman2016n1,Harshman2016n2,Dehkharghani2016,Barfknecht2016}, and the many-body Hamiltonian can be rewritten in terms of a center-of-mass (CM) coordinate $R=\frac{1}{\sqrt{N}}\sum_{n=1}^N x_n$ and $N-1$ relative Jacobi-coordinates (REL) given by $y_n=\sqrt{\frac{n}{n+1}}\left[ \frac{1}{n} \sum_{j=1}^n x_j-x_{n+1}\right]$. The system is then separable as $\hat{H}=\hat{H}_{\text{CM}}+\hat{H}_{\text{REL}}$, where the center-of-mass Hamiltonian corresponds to a single particle harmonic oscillator with frequency $\Omega(t)$, while the relative Hamiltonian contains the effects of interactions and is given by
\begin{align}
&\hat{H}_{\text{REL}} = \sum_{j=1}^{N-1} \left[ -\frac{1}{2}\frac{\partial^2}{\partial \hat{y}_j^2}+\frac{1}{2}  \Omega^2(t) \hat{y}_j^2 \right] + \nonumber\\
&\sum_{k>j} g \, \delta \left(\sqrt{\frac{j-1}{j}}\hat{y}_{j-1}-\sqrt{\frac{k-1}{k}}\hat{y}_{k-1}-\sum_{n=j}^k \frac{1}{\sqrt{n(n+1)}}\hat{y}_n\right) 
\label{eq:relativeH}
\end{align}
Rewriting the lab-frame position operators as $\hat{x}_n=\frac{\hat{R}}{\sqrt{N}}+\hat{Y}_n$, where $\hat{Y}_n=\sum_{j=n}^{N-1}{\frac{1}{\sqrt{j(j+1)}}\hat{y}_j}-\sqrt{\frac{n-1}{n}}\hat{y}_{n-1}$ is the collective relative coordinate, allows us to recast the infinite time-average as \cite{SuppMat} $\bar{C}_{\hat{x}_j, \hat{x}_k} =\bar{C}_{\hat{Y}_j,\hat{Y}_k}+\frac{1}{N^2}\bar{C}_{\hat{R},\hat{R}}$. The infinite time-average is therefore given simply as the sum of the CM and REL averages and this also holds when considering the momentum operators $\bar{C}_{\hat{x}_j, \hat{p}_k}$ and $\bar{C}_{\hat{p}_j, \hat{p}_k}$ \cite{SuppMat}.

For the CM coordinates the squared commutator is equivalent to the non-interacting system and given by $\bar{C}_{\hat{R},\hat{R}}=\frac{1}{2}$ \cite{Hashimoto2017,SuppMat}. This contribution to the full scrambling decreases with the system size as $\frac{1}{2N^2}$. The average scrambling in the system after a quench is therefore entirely determined by the relative-coordinate sector for which state-dependent energy-shifts for the even parity states resulting from the interaction ensure that condition (\textit{ii}) is obeyed. As the Hamiltonian has a full reflection symmetry with respect to the Jacobi coordinates $\hat{y}_n$ \cite{Harshman2016n2} one can prove condition (\textit{iii}) for the individual operators and therefore the squared commutator for $\hat{Y}_n$ fulfills conditions (\textit{i})-(\textit{iii})) and has an infinite time-average given by Eq.(\ref{eq:timeaverageofIandD}) \cite{SuppMat}. 

In order to show that $\bar{C}_{\hat{x}_1, \hat{x}_1}$ is proportional to the work fluctuations, however, we also require knowledge of the emergent operator $K^{Y_1,Y_1}$. From the analytically solvable $N=2$ case we find that $K^{Y_1,Y_1}$ from Eq.~\eqref{eq:timeaverageofIandD} is approximately a tri-diagonal matrix with the largest contribution from the elements $K^{Y_1,Y_1}_{jj}$, $K^{Y_1,Y_1}_{j,j+1}$ and $K^{Y_1,Y_1}_{j,j-1}$ which scale with leading terms proportional to $(E^{\text{REL}}_j)^2$. The second moment of the work probability distribution is given as $\langle W_{\text{REL}}^2 \rangle = \sum_{j} |c_{j}|^2 (E^{\text{REL}}_{j}-E^I)^2$, while $\bar{I}_{Y_1,Y_1}\propto \sum_{j}|c_{j}|^2 (E^{\text{REL}}_{j})^2$ as it is only diagonal in $K^{Y_1,Y_1}_{jj}$. While this clearly links the dynamics of the correlation functions to the second moment of the work probability distribution, it also holds for its variance $\Delta W^2 = \sum_j |c_{j}|^2 (E^{\text{REL}}_{j})^2-\left(\sum_j|c_{j}|^2 E^{\text{REL}}_{j}\right)^2$  as $\langle W^2 \rangle \propto \langle W \rangle^2$. A similar argument can be made for $\bar{D}_{Y_1,Y_1}$ which can effectively be described as a sum over the tri-diagonal elements of $K^{Y_1,Y_1}$ \cite{SuppMat}. This final condition (\textit{iv}) [$K^{A,B}_{jj} \propto  (E^{\text{REL}}_{j})^2$] is required to link the scrambling to the work fluctuations. While this is satisfied for a harmonic trap one cannot expect this result to generalize to other systems. However, the scrambling in any system which obeys conditions (\textit{i})-(\textit{iii})) will be closely connected to the work statistics through Eq.\eqref{eq:timeaverageofIandD}, although the relation can be more complicated depending on the properties of $K^{A,B}$.

\begin{figure}
\centering
\includegraphics[width=0.999\linewidth]{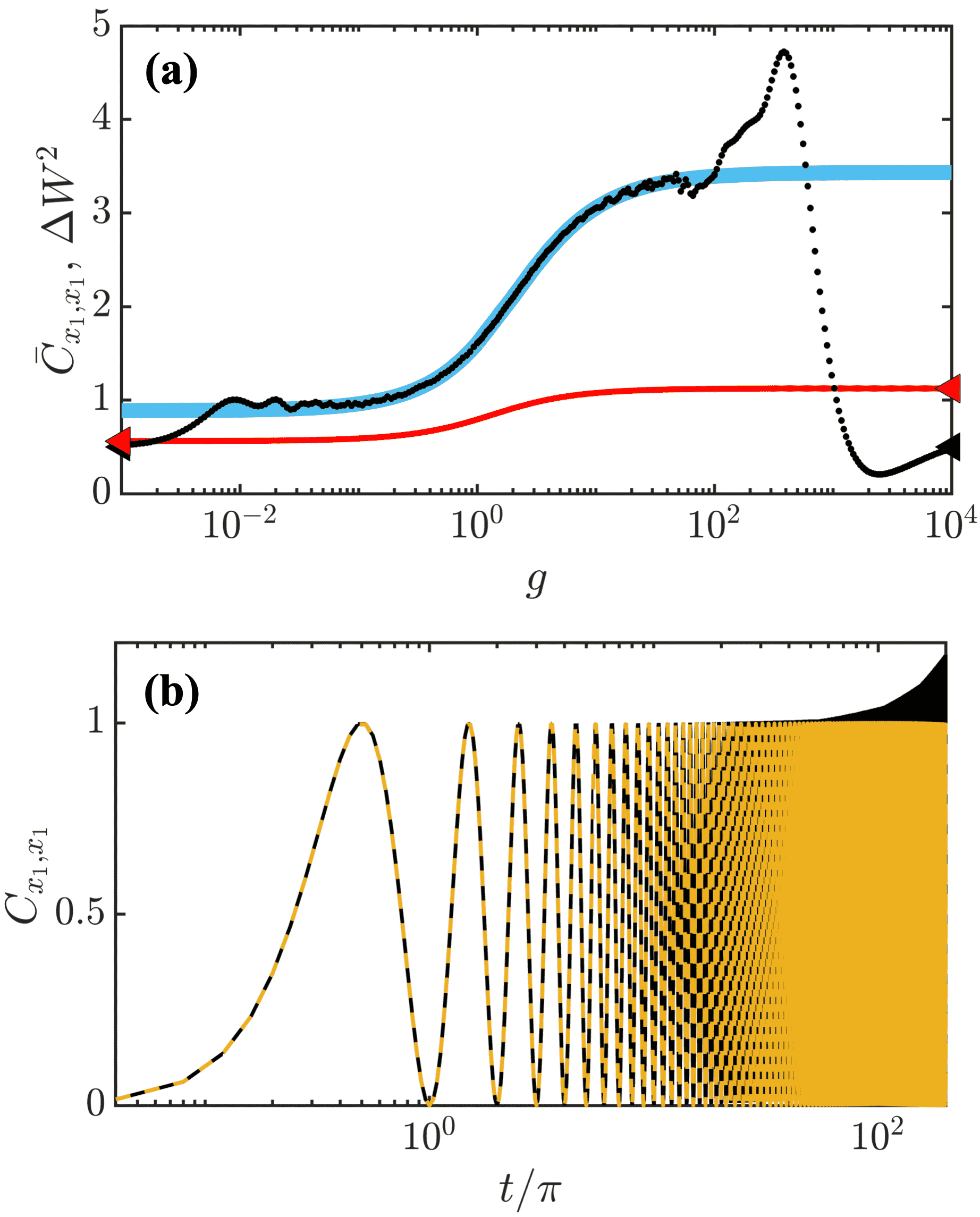}
\caption{(a) The full blue line is the infinite-time average of the squared commutator $\bar{C}_{x_1,x_1}$ as a function of $g$ for  $\gamma = \frac{1}{2}$. Black dots show the time-average of the same squared commutator in an interval $t\in [0,200 \pi ]$ as a function of $g$. The variance of the work distribution function $\Delta W^2$ is given by the red solid line. The triangles correspond to the $g=0$ and $g=\infty$ values of $\bar{C}_{xx}$ (black) and $\Delta W^2$ (red).
(b) $C_{x_1,x_1}(t)$ as a function of time for $g =0.002$ (black) and $g=0$ (yellow dashed) with $\gamma=\frac{1}{2}$.}
\label{fig:Timeaverage1} 
\end{figure}

As noted previously, there is no connection between the variance and the information scrambling when the particles are in the limits of zero and infinite interactions, however, for finite interactions a linear relationship was found. Next we explore how this manifests as a function of the interaction strength for $N=2$. Using Eq.~\eqref{eq:timeaverageofIandD} the infinite-time average of the squared commutator $\bar{C}_{\hat{x}_1,\hat{x}_1}$ can be calculated as a function of the interaction $g$, which is shown in Fig.~\ref{fig:Timeaverage1}(a). The information scrambling increases with increasing interactions and reaches asymptotic values for $g\rightarrow 0$ and $g\rightarrow \infty$. However, these asymptotic values are different from the known values in the limits $g=\{0,\infty \}$, given as $\bar{C}_{\hat{x}_1,\hat{x}_1}=1/2$ (black triangles in the figure). In contrast the work fluctuations show a smooth crossover to the limiting values (red lines and triangles respectively).

The difference between the asymptotic values of the squared commutator and the limiting values at $g=\{0,\infty \}$ shows that the scrambling is very sensitive to small deviations from the harmonic oscillator spectrum on infinitely long timescales. To check this result we compute the full time-dependent OTOCs in Eqs.~(\ref{eq:Dgenericexplicit}-\ref{eq:Fgenericexplicit}) and numerically find their time average in the range $t\in [0,200 \pi]$ (black dots in Fig.~\ref{fig:Timeaverage1}(a)). For intermediate values of $g \in[0.1,70]$ these results are indistinguishable from each other, but as the extremal interaction limits are approached the finite time-average and infinite time-average results diverge. For the $g=0$ limit the dynamics are shown in Fig.~\ref{fig:Timeaverage1}(b) and are simply given by $C_{\hat{x}_1,\hat{x}_1}(t)=\sin^2(t)$ (yellow dashed line). It is interesting to compare them to the case of weak interactions, $g=0.002$, which possesses equivalent dynamics on short timescales (black solid line). In this case the interaction induced energy shift $\Delta_j$ is small and decreases as $\Delta_j \propto j^{-1/2}$ \cite{Busch98}, such that the dynamics on short times can be approximated as $e^{-i (E_j^{0}+\Delta_j)t}\approx e^{-i (E_j^{0})t}$ with $E_j^{0}$ being the single particle harmonic oscillator energies. However at long times these energy shifts will affect the dynamics, leading to a change in the time average that is captured by Eq.~\eqref{eq:timeaverageofIandD}. This discontinuity in the average information scrambling is therefore only observable in the long-time limit as the timescale required to observe the average scrambling diverges (similar for the $g\rightarrow \infty$ case).

In summary, we have shown that for harmonically trapped interacting atoms, which are a fundamental building block in many cold atom experiments, the time-average of the
squared commutator $C_{AB}(t)$ for canonical operators is proportional to the work fluctuations. The operator scrambling in Hilbert space is therefore intimately linked to the work probability distribution, which is an experimentally accessible thermodynamic measure \cite{Fusco2014,Serra2014,Cerisola2017} of the non-equilibrium excitations induced by the quench. However, the timescale required to observe information scrambling is interaction dependent, being shorter the further the system is from the harmonic limits. In fact it diverges as the non-interacting and TG limits are approached, highlighting the importance of intermediate interactions to be able to observe information scrambling on short time-scales. The relative lack of finite-size effects is curious and a further investigation of the moments of the work probability distribution as a function of $g$ and $N$ in a harmonic trap is an interesting line for future investigations. It would also be interesting to investigate other potentials which obey conditions (\textit{i})-(\textit{iii}), but likely not condition (\textit{iv}), in order to contrast and compare with the case of harmonic trapping.

\begin{acknowledgments}
\section{Aknowledgements}
This work was supported by  the  Okinawa  Institute  of  Science  and  Technology  Graduate  University and utilized the computing resources of the Scientific Computing and Data Analysis section of the Research Support Division at OIST. In addition, MM was supported by the Japan Society for the Promotion through the JSPS  fellowship  (JSPS  KAKENHI  Grant No. 19J10852). TF acknowledges support under JSPS KAKENHI - 21K13856.
\end{acknowledgments}

\FloatBarrier

\newpage


\section*{Supplementary material}

# Supplemental Material

## DIAGONAL EXPECTATION VALUES OF CANONICAL OPERATORS FOR SYSTEMS WITH FULL REFLECTION SYMMETRY

An arbitrary bosonic many-body quantum state can be expanded in a product-basis of many-body wavefunctions

$$\Psi_{x_1,...,x_N} = \mathcal{N} \sum_\nu c_\nu \sum_{\sigma \in S_n} \prod_n \psi_{\nu_{\sigma(n)}}(x_n), \quad (S.1)$$

where the $\psi_{\nu_{\sigma(n)}}(x_n)$ form a single-particle basis. Here the $S_n$ are all the permutations of $1, 2, ..., N$ and $\mathcal{N}$ is the normalisation factor.

We now consider a system which obeys a full reflection symmetry with respect to its coordinates, i.e. with eigenstates which obey $\Psi_n(x_1, ..., x_N) = \pm\Psi_n(-x_1, ..., -x_N)$, of which one example are interacting particles in a harmonic trap.

Due to the reflection symmetry states are either even or odd and the two sectors completely decouple. An arbitrary many-body wavefunction can then be written as

$$\Psi_{x_1,...,x_N} = \mathcal{N} \sum_\nu c_\nu \sum_{\sigma \in P_n} \prod_m \psi_{\nu_{\sigma(m)}}(x_m), \quad (S.2)$$

where $P_n$ are all permutations that have an even number of odd single-particle wavefunctions in the even sector or all permutations that have an odd number of odd single-particle wavefuntions in the odd sector. The diagonal expectation value of a canonical one-body operator with respect to the many-body states is then given by ($A_j = x_j, p_j$)

$$\langle A_j \rangle = \mathcal{N}^2 \sum_{\sigma \in P_n} \int dx_1...dx_N \sum_\nu c_\nu^* \prod_m \left(\psi_{\nu_{\sigma(m)}}\right)^*(x_m) \times$$
$$A_j \sum_{\sigma \in P_n} \sum_\mu c_\mu \prod_m \psi_{\mu_{\sigma(m)}}(x_m). \quad (S.3)$$

We now consider an arbitrary term in the sum given by

$$c_\nu^* c_\mu \langle \psi_{\nu_{\sigma(j)}}^{\text{odd/even}}(x_j) | A_j | \psi_{\mu_{\sigma(j)}}^{\text{odd/even}}(x_j) \rangle \times$$
$$\prod_{m \neq j} \langle \psi_{\nu_{\sigma(m)}}(x_m) \prod_{m \neq j} | \psi_{\mu_{\sigma(m)}}(x_m) \rangle. \quad (S.4)$$

If the operator connects two even or two odd states the resulting integral will be over an odd function as $x_j$ is an odd function and $p_j$ can be written as a differential operator which changes the parity of even and odd functions. The integral is therefore zero. If the operator connects an odd and an even SP state on the other hand this term is not zero and we have to consider the remaining inner products. In this case one odd or even function has been removed from the left-hand side and correspondingly one even or odd function has been removed from the right-hand side. This means that the number of odd and even states on the left and right-hand side is unequal and an inner product between an odd and even single-particle state, which gives zero, is always present. All terms in the expectation value therefore gives zero and the expectation value itself vanishes.

## EVALUATING THE LAB-FRAME SQUARED COMMUTATORS FOR N PARTICLES

The Hamiltonian in Eq.(4) in the main text for $N$ interacting particles can be separated into a center-of-mass (CM) part and a relative part. The full state can therefore be written as a product of the eigenstates of these two Hamiltonians. The CM eigenstates $|\psi_n^{\text{CM}}\rangle$ correspond to those of a non-interacting harmonic oscillator while the relative eigenstates $|\psi_\nu^{\text{REL}}\rangle$ diagonalize Eq.(13) in the main text and carries all information about the interaction. The initial state after a quench can therefore be written in terms of the eigenstates of the final Hamiltonian (Eq.(4) in the main text) as $|\psi_I^{\text{CM}}\rangle|\psi_I^{\text{REL}}\rangle = \sum_{\nu,n} d_{2n+1} c_\nu |\psi_{2n+1}^{\text{CM}}\rangle|\psi_\nu^{\text{REL}}\rangle$ for odd parity CM states and $|\psi_I^{\text{CM}}\rangle|\psi_I^{\text{REL}}\rangle = \sum_{\nu,n} d_{2n} c_\nu |\psi_{2n}^{\text{CM}}\rangle|\psi_\nu^{\text{REL}}\rangle$ for even parity CM states.

These sums only involve even or odd CM eigenstates because these symmetry sectors of the Hamiltonian are decoupled, which means that any quench for which the CM initial state is even or odd only involves even or odd states in the dynamics. Similarly the dynamics for the relative part only involve states in the same symmetry sector as the initial state, which means that relative states with different parity have zero overlap. The overlap coefficients are given by $c_\nu = \langle \psi_\nu^{\text{REL}} | \psi_I^{\text{REL}} \rangle$ and $d_n = \langle \psi_n | \psi_I^{\text{CM}} \rangle$. Utilizing this representation of the state the correlation functions that go into the squared commutator for the canonical operators can be calculated. As an example we consider $I_{x_j,x_k}(t)$ which is then given by (for an even initial CM state, the derivation is identical for an odd initial CM state)

$$I_{x_j,x_k}(t) = \sum_{\nu_1, n_1, \nu_2, n_2} c_{\nu_1} d_{2n_1} c_{\nu_2} d_{2n_2} \langle \psi_{\nu_1}^{\text{REL}} | \langle \psi_{2n_1}^{\text{CM}} |$$
$$[\hat{Y}_j(t) + \frac{\hat{R}(t)}{\sqrt{N}}][\hat{Y}_k + \frac{\hat{R}}{\sqrt{N}}]^2 [\hat{Y}_j(t) + \frac{\hat{R}(t)}{\sqrt{N}}] | \psi_{2n_2}^{\text{CM}} \rangle | \psi_{\nu_2}^{\text{REL}} \rangle. \quad (S.5)$$

Here $\hat{R}$ and $\hat{Y}_j$ are the CM and collective relative coordinates defined in the main text. Any term which involves an odd number of CM coordinate operators will be zero

due to the structure of the canonical operators. This is because such a combination results in an odd parity state when applied to the even parity states, which will have zero overlap with the even parity states contained in the sum. The correlation functions can therefore be expressed as (where $K = I, D$)

$$K_{A_j B_k}(t) = K_{A_j^{\text{REL}} B_k^{\text{REL}}}(t) + \frac{K_{A^{\text{CM}} B^{\text{CM}}}(t)}{N^2}$$
$$+ \frac{1}{N}\bigg(J_{B^{\text{CM}}}(0) J_{A_j^{\text{REL}}}(t) + J_{B_k^{\text{REL}}}(0) J_{A^{\text{CM}}}(t)$$
$$+ 2[G^1_{A_j^{\text{REL}} B_k^{\text{REL}}}(t) G^2_{A^{\text{CM}} B^{\text{CM}}}(t)$$
$$+ G^2_{A_j^{\text{REL}} B_k^{\text{REL}}}(t) G^1_{A^{\text{CM}} B^{\text{CM}}}(t)]\bigg), \quad (S.6)$$

$$F_{A_j B_k}(t) = F_{A_j^{\text{REL}} B_k^{\text{REL}}}(t) + \frac{F_{A^{\text{CM}} B^{\text{CM}}}(t)}{N^2}$$
$$+ \frac{1}{N}\bigg(J_{B^{\text{CM}}}(0) J_{A_j^{\text{REL}}}(t) + J_{B_k^{\text{REL}}}(0) J_{A^{\text{CM}}}(t)$$
$$+ G^1_{A_j^{\text{REL}} B_k^{\text{REL}}}(t) G^2_{A^{\text{CM}} B^{\text{CM}}}(t)$$
$$+ G^2_{A_j^{\text{REL}} B_k^{\text{REL}}}(t) G^1_{A^{\text{CM}} B^{\text{CM}}}(t)$$
$$+ 2 G^1_{A_j^{\text{REL}} B_k^{\text{REL}}}(t) G^1_{A^{\text{CM}} B^{\text{CM}}}(t)\bigg), \quad (S.7)$$

where two-point correlation functions (2-PCF) are defined as

$$G^1_{AB} = \langle A(t) B \rangle \quad, \quad G^2_{AB} = \langle B A(t) \rangle \quad, \quad J_A(t) = \langle A^2(t) \rangle. \quad (S.8)$$

The expectation values are with respect to the relative wavefunction $|\psi_I^{\text{REL}}\rangle$ for relative coordinate operators and the CM wavefunction $|\psi_I^{\text{CM}}\rangle$ for CM coordinate operators. The full squared commutator of the laboratory-frame coordinates is given by

$$C_{A_j, B_j} = C_{A_j^{\text{REL}} B_k^{\text{REL}}}(t) + \frac{C_{A^{\text{CM}} B^{\text{CM}}}(t)}{N^2}$$
$$+ \frac{2}{N}[G^1_{A_j^{\text{REL}} B_k^{\text{REL}}}(t) G^2_{A^{\text{CM}} B^{\text{CM}}}(t)$$
$$+ G^2_{A_j^{\text{REL}} B_k^{\text{REL}}}(t) G^1_{A^{\text{CM}} B^{\text{CM}}}(t)$$
$$- 2 G^1_{A_j^{\text{REL}} B_k^{\text{REL}}}(t) G^1_{A^{\text{CM}} B^{\text{CM}}}(t)]. \quad (S.9)$$

The scrambling is therefore given in terms of the squared commutator expectation values of the center of mass and relative coordinates, plus products of 2-PCF. The later are related to the unequal time commutator rather than the unequal time squared commutator.

An example of the explicit time-dependence of these 2-PCF products is given by

$$G^1_{A_j^{\text{REL}} B_k^{\text{REL}}}(t) G^1_{A^{\text{CM}} B^{\text{CM}}}(t) =$$
$$\sum_{\nu\nu'\mu n n' m} c_\nu^* c_{\nu'} d_{2n}^* d_{2n'} e^{i(E_\nu^{\text{REL}} - E_\mu^{\text{REL}} + E_{2n}^{\text{CM}} - E_m^{\text{CM}})t} \times$$
$$\left(A_j^{\text{REL}}\right)_{\nu,\mu} \left(B_k^{\text{REL}}\right)_{\mu,\nu'} A^{\text{CM}}_{2n,m} B^{\text{CM}}_{m,2n'}. \quad (S.10)$$

In the interacting system, the energy shifts induced by the delta-interaction means that $E_\nu^{\text{REL}} - E_\mu^{\text{REL}} + E_{2n}^{\text{CM}} - E_m^{\text{CM}} = 0$ only holds when $E_\nu^{\text{REL}} - E_\mu^{\text{REL}} = 0$ and $E_{2n}^{\text{CM}} - E_m^{\text{CM}} = 0$. These two conditions are in turn only fulfilled for $\mu = \nu$ and $m = 2n$ due to the non-degeneracy of the two spectra. This means that the infinite time-average is given by

$$\sum_{\nu\nu' n n'} c_\nu^* c_{\nu'} d_{2n}^* d_{2n'} \left(A_j^{\text{REL}}\right)_{\nu,\nu} \left(B_k^{\text{REL}}\right)_{\nu,\nu'} A^{\text{CM}}_{2n,2n} B^{\text{CM}}_{2n,2n'} \quad (S.11)$$

which is equal to zero as the system fulfills condition $(iii)$ in the main text ($B_{nn} = A_{nn} = 0$). This also holds true for the other 2-PCF products. The infinite time-average of the lab-frame coordinate scrambling is therefore given as the sum of the relative and center of mass squared commutator infinite time averages

$$\bar{C}_{A_j, B_k} = \bar{C}_{A_j^{\text{REL}}, B_k^{\text{REL}}} + \frac{\bar{C}_{A^{\text{CM}}, B^{\text{CM}}}}{N^2}. \quad (S.12)$$

Note that for the non-interacting system the 2-PCF products don't average to zero and have to be taken into account to match the solutions obtained by directly considering the lab-frame coordinates.

### SQUARED COMMUTATOR FOR $g = 0, \infty$ IN THE HARMONIC TRAP

For a non interacting system of $N$ particles the Hamiltonian is given by the sum of single-particle Hamiltonians $\hat{H} = \sum_{j=1}^N \hat{H}_j$, where $[\hat{H}_j, \hat{H}_k] = 0$. The time-evolution operator can then be written as $U = \prod_{j=1}^N e^{-i\hat{H}_j t}$ which means that the time-evolution of any canonical operator is given by $\hat{A}_j = e^{i\hat{H}_j t} \hat{A}_j e^{-i\hat{H}_j t}$. Combined with $[\hat{A}_j, \hat{B}_k] = 0$ and $[\hat{H}_j, \hat{B}_k] = 0$ it is clear that $[\hat{A}_j(t), \hat{B}_k] = 0$ for $j \neq k$. The time-dependent squared commutator will therefore only be finite for $j = k$. Since for $N$ non-interacting bosons, the initial state for a trap quench is given by

$$\psi^I(x_1, x_2, ..., x_N) = \prod_n \psi_0(x_n) \quad (S.13)$$

where $\psi_0(x_n)$ is the ground-state of the corresponding single-particle Hamiltonian, the squared commutator for

$j = k$ is given by

$$\langle [\hat{A}_j(t), \hat{B}_j]^2 \rangle = \prod_n \langle \psi_0(x_n)|[\hat{A}_j(t), \hat{B}_j]^2 \prod_m |\psi_0(x_m)\rangle$$
$$= \langle \psi_0(x_j)|[\hat{A}_j(t), \hat{B}_j]^2|\psi_0(x_j)\rangle. \quad \text{(S.14)}$$

The squared commutator of canonical operators is therefore given by the single-particle result as expected and is independent of system-size.

In the TG limit of strongly interacting bosons the system can be described by the non-interacting Hamiltonian $\hat{H} = \sum_{j=1}^{N} \hat{H}_j$ with the interactions taken into account by the following constraint on the many-body wavefunction: $\psi^I(x_1, x_2, ..., x_N) = 0$ when $x_j - x_k = 0$. This constraint is fulfilled by the equivalent fermionic wave-function which has to be multiplied with the antisymmetric unit function to fulfill the bosonic symmetrization requirements [1]

$$\psi^I(x_1, x_2, ..., x_N) = \prod_{j<k} \text{sgn}(x_k - x_j)\psi_F(x_1, x_2, ..., x_N). \quad \text{(S.15)}$$

Here the fermionic wavefunction is given by the Slater determinant

$$\psi_F(\vec{x}) = \frac{1}{\sqrt{N!}}\det(S) \quad , \quad S_{jn} = \psi_j(x_n) \quad \text{(S.16)}$$

where $\vec{x} = \{x_1, x_2, ..., x_N\}$. The time-dependent commutator for $j \neq k$ is given by $[\hat{A}_j(t), \hat{B}_k] = 0$ using the same argument as for the free Bose gas (as the time-evolution operator is the same) and we therefore focus on the case $j = k$.

We will first show that the TG gas has the same value of the squared commutator as the free Fermi gas. This holds if the squared commutator commutes with the sign function in which case the sign functions will cancel out when taking the expectation value of the squared commutator. In order to show this we must show that it commutes with the individual operators for which we take the squared commutator as well as the time-evolution operator. By writing the time-evolution operator of the non-interacting Hamiltonian in terms of a power-series this means that the sign function has to commute with an arbitrary product of position and momentum operators (which also shows that it commutes with the individual operators in the squared commutator).

It is immediately clear that any power of any position coordinate (including $m = j, k$) commutes with any sign function $[x_m^p, \text{sgn}(x_k - x_j)] = 0$. Since applying the momentum operator amounts to taking the derivative with respect to that coordinate, any $\text{sgn}(x_k - x_j)$ factors for which $m \neq j, k$ trivially commute with the derivative as well and we therefore focus on the factors where $m = j$ or $m = k$,

$$\frac{\partial}{\partial x_m}\left(\text{sgn}(x_k - x_j)\psi_F(\vec{x})\right) =$$
$$\text{sgn}(x_k - x_j)\frac{\partial \psi_F(\vec{x})}{\partial x_m} + \psi_F(\vec{x})\frac{\partial(\text{sgn}(x_k - x_j))}{\partial x_m}. \quad \text{(S.17)}$$

We observe that $\frac{\partial(\text{sgn}(x_k - x_j))}{dx_m} = 0$ for $x_k - x_j \neq 0$ while $\psi_F(\vec{x}) = 0$ for $x_k - x_j = 0$. The derivative is therefore equal to $\text{sgn}(x_k - x_j)\frac{\partial \psi_F(\vec{x})}{\partial x_m}$. Higher powers of the momentum operator will behave similarly, as $\frac{\partial^n \psi_F(\vec{x})}{\partial x_m^n} = 0$ for $x_k - x_j = 0$. This can be understood by applying Jacobi's rule for derivatives of determinants which implies $\frac{\partial(\det(S))}{\partial x_m} = \det(S)\text{Tr}(S^{-1}\frac{\partial S}{\partial x_m})$ with the chain rule ensuring that all terms for higher derivatives will involve the factor $\det(S) = \sqrt{N!}\psi_F(\vec{x}) = 0$ for $x_k - x_j = 0$. This also ensures that any products of momentum and position operators commute with the sign function. The only change to the above calculation is that one considers the derivative of $x_m^p \psi_F(\vec{x})$ with the extra term disappearing by the same argument.

For $j = k$ the squared commutator is therefore equivalent to that of a free Fermi gas and given by

$$\langle [\hat{A}_j(t), \hat{B}_j]^2 \rangle = \frac{1}{N!}\det S[\hat{A}_j(t), \hat{B}_j]^2 \det S$$
$$= \frac{1}{N}\sum_{n=1}^{N}|\langle \psi_n(x_j)^*|[\hat{A}_j(t), \hat{B}_j]^2|\psi_n(x_j)\rangle \quad \text{(S.18)}$$

i.e. a sum of the squared commutator from each eigenstate below the Fermi energy.

To get the explicit final result, the single-particle squared commutators are required. We utilize the same scaled units in this section as in the main text. This means that all times are given in terms of the trap frequency and all length scales are given in terms of the natural length of the final Hamiltonian. For a non-interacting particle in a harmonic trap the full Heisenberg equations of motion for position and momentum operators are given by

$$\hat{x}(t) = \hat{x}(0)\cos(t) + \hat{p}(0)\sin(t) \quad \text{(S.19)}$$
$$\hat{p}(t) = \hat{p}(0)\cos(t) - \hat{x}(0)\sin(t). \quad \text{(S.20)}$$

This can be utilized to calculate the time-dependent correlation functions by simple multiplication and taking the expectation-value of the time-independent correlation functions with respect to the initial state. The squared commutator for an eigenstate of this system has been calculated in earlier works, such as [2]. Here we consider the trap quench and find that $I_{AB}(t) = D_{AB}(t)$

with

$$I_{AA}(t) = \frac{3}{4}[G_A \cos^2(t) + \sin^2(t)], \quad (S.21)$$

$$\text{Re}\,[F_{AA}(t)] = \frac{3}{4}[G_A \cos^2(t) + \sin^2(t)] - \frac{1}{2}\sin^2(\omega_F t), \quad (S.22)$$

$$I_{AB}(t) = \frac{3}{4}[\cos^2(t) + G_A \sin^2(t)], \quad (S.23)$$

$$\text{Re}\,[F_{AB}(t)] = \frac{3}{4}[\cos^2(t) + G_A \sin^2(t)] - \frac{1}{2}\cos^2(t), \quad (S.24)$$

with the squared commutator given by

$$C_{AA}(t) = \sin^2(t), \quad (S.25)$$
$$C_{AB}(t) = \cos^2(t). \quad (S.26)$$

Here $G_x = \frac{1}{\gamma^2}$ and $G_p = \gamma^2$, leading to an operator-dependent difference between squeezing and opening the trap. Similar formulas can easily be found for initially excited states. While the individual correlation functions become more complicated, the squared commutator is always given by Eqs.(S.25) and (S.26) and it is therefore independent of the initial state. This means that in the non-interacting and the TG-limit the squared commutator of lab frame coordinates is given by Eqs.(S.25,S.26) for $j = k$, while it is zero otherwise.

### TIME-DEPENDENT RELATIVE COORDINATE CORRELATION FUNCTIONS FOR $N = 2$

For two particles the relative Hamiltonian is given by

$$\hat{H}_{\text{rel}} = -\frac{1}{2}\frac{\partial^2}{\partial x^2} + \frac{1}{2}\Omega^2 x^2 + g\,\delta(x). \quad (S.27)$$

The problem is then analytically solvable with eigenenergies for the even parity eigenstates given by the transcendental equation [3]

$$\frac{\Gamma(-\frac{E_{2j}^{\text{REL}}}{2} + \frac{3}{4})}{\Gamma(-\frac{E_{2j}^{\text{REL}}}{2} + \frac{1}{4})} = -\frac{g}{2}, \quad (S.28)$$

while the odd parity eigenstates are not affected by the interaction and are given by the pure harmonic oscillator ones. The energy of the $2j^{\text{th}}$ state can also be expressed as

$$E_{2j}^{\text{REL}} = (\frac{1}{2} + 2j + \Delta_j)\Omega \quad (S.29)$$

A convenient representation of the even parity relative wave-functions in the presence of interactions is given by

[4]

$$\psi_{2j}^{\text{REL}}(x) = A_{2j}^g \sum_{0 \leq n \leq \infty} \frac{\psi_{2n}^*(0)}{E_{2n} - E_{2j}^{\text{REL}}} \psi_{2n}(x), \quad (S.30)$$

where $E_{2n}$ is the energy of the $2n^{\text{th}}$ non-interacting HO eigenstate $\psi_{2n}$. The normalization factor is given by

$$A_{2j}^g = \sqrt{\frac{4\Gamma(-\frac{E_{2j}^{\text{REL}}}{2} + \frac{3}{4})}{\Gamma(-\frac{E_{2j}^{\text{REL}}}{2} + \frac{1}{4})[\tilde{\Gamma}(-\frac{E_{2j}^{\text{REL}}}{2} + \frac{3}{4})) - \tilde{\Gamma}(-\frac{E_{2j}^{\text{REL}}}{2} + \frac{1}{4})]}}, \quad (S.31)$$

where $\Gamma$ and $\tilde{\Gamma}$ are the gamma and di-gamma functions respectively.

This representation allows us to calculate the effect of position and momentum operators ($\hat{O} = \hat{x}, \hat{p}$) on the interacting state as their action on the non-interacting harmonic oscillator states $|\psi_n\rangle$ are known to be

$$\hat{O}|\psi_n\rangle = \frac{\beta_O}{\sqrt{2}}\left(\sqrt{n+1}|\psi_{n+1}\rangle + \tilde{\beta}_O\sqrt{n}|\psi_{n-1}\rangle\right), \quad (S.32)$$

with $\beta_O = 1$, $\tilde{\beta}_O = 1$ for the position operator and $\beta_O = i$, $\tilde{\beta}_O = -1$ for the momentum operator. Here we assume $\Omega = 1$ corresponding to the the final Hamiltonian for the quench considered in the main text. The correlation functions for a quench in the interacting system, assuming a quench from an even parity initial state $|\psi_I\rangle$ and real coefficients $c_{2j} = \langle \psi_{2j}^g | \psi_I \rangle$ can then be found, utilizing Eq.(S.30) as

$$D_{AB}(t) = \sum_{j,k} c_{2j} c_{2k} K_{jk}^{BA} + \cos(2t) \sum_{j,k} c_{2j} c_{2k} J_{jk}^{BA}, \quad (S.33)$$

$$I_{AB}(t) = \sum_j |c_{2j}|^2 [K_{jj}^{AB} + J_{jj}^{AB}\cos(2t)]$$
$$+ \sum_{j \neq k} c_{2j}c_{2k}\left(K_{jk}^{AB}\cos[(E_{2j}^g - E_{2k}^g)t] + J_{jk}^{AB}\cos[(E_{2j}^g - E_{2k}^g + 2)t]\right), \quad (S.34)$$

$$F_{AB}(t) = \sum_{j,k,n,m,l} c_{2j}c_{2k} e^{-i(E_{2n}^g - E_{2k}^g + 2l - 2m)t} \times \alpha_{m,n}^B \alpha_{m,k}^A \alpha_{l,n}^A \alpha_{l,j}^B. \quad (S.35)$$

Here

$$K_{jk}^{AB} = \sum_{m=0}^{\infty} \frac{|\beta_A|^2 |\beta_B|^2 \tilde{\beta}_B}{2} \alpha_{m,j}^A \alpha_{m,k}^A (4m+3), \quad (S.36)$$

$$J_{jk}^{AB} = \sum_{n=1}^{\infty} |\beta_A|^2 |\beta_B|^2 \sqrt{2n+1}\sqrt{2n}\,\alpha_{n,k}^A \alpha_{n-1,j}^A, \quad (S.37)$$

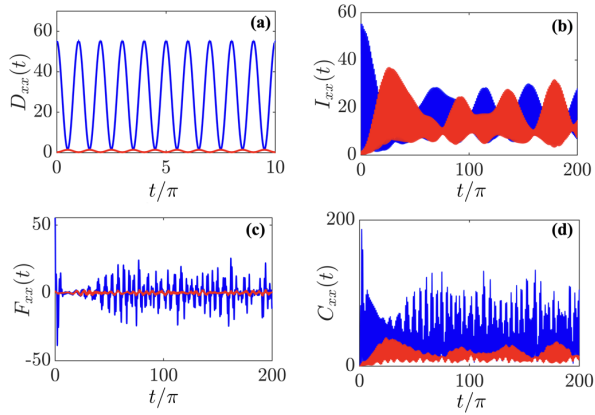

FIG. 1. $D_{xx}(t)$ (a), $I_{xx}(t)$ (b), $F_{xx}(t)$ (c) and $C_{xx}(t)$ (d) as a function of time for $g = 10$. The blue lines correspond to a quench with $\gamma = \frac{1}{4}$, while the red lines correspond to a quench with $\gamma = 4$.

and

$$\alpha^O_{m,j} = \langle \psi_{2m+1} | \hat{O} | \psi^{\text{REL}}_{2j} \rangle / \beta_O$$
$$= \sqrt{2m+1} \psi_{2m}(0) \frac{A_{2j}}{\sqrt{2}} \left( \frac{1}{E_{2m} - E^{\text{REL}}_{2j}} - \frac{\tilde{\beta}_O}{E_{2m+2} - E^{\text{REL}}_{2j}} \right). \quad \text{(S.38)}$$

In the latter the structure of the position/momentum operator means that only odd parity non-interacting states remain when applied to the sum of even parity non-interacting states given by Eq.(S.30), which is why these coefficients are only non-zero for $|\psi_{2m+1}\rangle$. We note that the time-ordered correlation function $D_{AB}(t)$ has an analytically simple time-dependence given by oscillations at twice the trapping frequency of the final Hamiltonian, as also shown in Fig. 1. A more thorough investigation of the time-dependence of the squared commutator is not the main focus of this work, but in Fig. 1 we show an example of the dynamics for opening and squeezing the trap. For these finite interactions the harmonicity of the energy spectrum is broken and therefore the correlation functions can possess complex dynamics. Indeed, $I_{xx}(t)$ and $F_{xx}(t)$ possess irregular oscillations as they are not time-ordered. It is clear that the time average of $F_{xx}(t)$ quickly vanishes, with the long time behaviour of the squared commutator determined solely by $D_{xx}(t)$ and $I_{xx}(t)$.

## ANALYSIS OF THE MATRIX ELEMENTS $K_{jk}$ FOR $N = 2$

The important coefficients that determine the behavior of the emergent matrix elements are given by Eq. (S.38). Let us first consider what happens for large $m, j$. To do this we establish how the quantities in Eq.(S.38) behave. The harmonic oscillator wavefunctions evaluated at zero argument $\psi_{2n}(0)$ are given by

$$\psi_{2m}(0) = \left( \frac{-1}{2} \right)^m \left( \frac{1}{\pi} \right)^{1/4} \frac{\sqrt{2m!}}{m!}, \quad \text{(S.39)}$$

which, by utilizing Stirling's formula, can be evaluated for large $m$ as

$$\psi_{2m}(0) \approx (-1)^m m^{-1/4} \propto m^{-1/4}. \quad \text{(S.40)}$$

The behaviour of the digamma functions for solutions of Eq. (S.28) is complicated, but numerically we find that $A_{2j} \propto j^{-1/4}$ for large $j$. The energy of the $2j$-th state can be expressed as in Eq. (S.29). For large $j$ one has $\Delta_j \propto j^{-1/2}$ (see [3]). Let us first investigate the convergence of $K_{jk}$ with respect to $m$. For $K_{jk}$, which involves the momentum-coefficients $\alpha^p_{m,j}$ the terms scale $\propto \psi^2_{2m}(0) = m^{-1/2}$, which means that the sum diverges. Numerical calculations for a short-range Gaussian interaction in the next section show that this divergence disappears in a physically realistic system. The sums can therefore be regularised by choosing a cut-off value $m_{\max}$ that is related to the non-universal finite-range parameter of the specific interaction considered. In order to evaluate a squared commutator involving a momentum-operator such a finite-range parameter is therefore required in addition to the zero-range interaction strength $g$. For $C_{xx}$, however, the coefficients can be rewritten as

$$\alpha^x_{m,j} = \sqrt{2m+1} \psi_{2m}(0) A_{2j} \frac{2}{(E_{2m} - E^{\text{REL}}_{2j})(E_{2m+2} - E^{\text{REL}}_{2j})}, \quad \text{(S.41)}$$

which means that the terms in $K_{jk}$ scale $\propto m^{-2} \psi^2_{2m}(0) = m^{-5/2}$, ensuring convergence.

Let us now consider the dependence of $\alpha^x_{m,j}$ on $j$. Using Eq. (S.29) we can rewrite the position coefficients as

$$\alpha^x_{m,j} = a_\omega \sqrt{2m+1} \psi_{2m}(0) A_{2j}$$
$$\times \left( \frac{1}{(2m - 2j + \Delta_j)} - \frac{1}{(2m + 2 - 2j + \Delta_j)} \right). \quad \text{(S.42)}$$

For $m = j$ and $m = j - 1$ the fractions scale $\propto j^{1/2}$ in the limit of large $j$, while the remaining terms do not scale with increasing j, which means that they become insignificant in comparison. As $m = j$ this means that the coefficients themselves scale as $\alpha^x_{j,m} \propto j^{1/2}$ in the limit of large $j$. In order to evaluate the long-time averages we need to investigate how this observation is manifested in the matrix-elements $K_{jk}$. These are only large when $j = m, j = m + 1$ and $k = m, k = m + 1$ is simultaneously true. This is only the case for $k = j, k = j - 1, k = j + 1$. So for large $j$ we expect that these values will be dominant, as the other matrix-elements are of insignificant

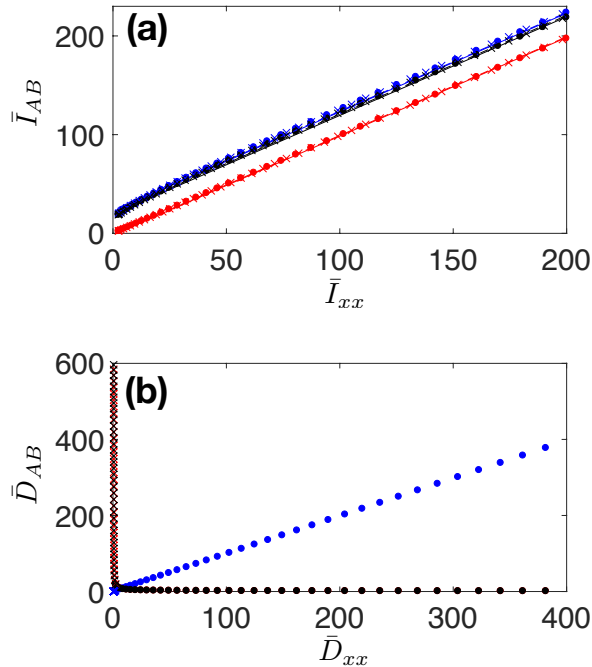

FIG. 2. (a) $\bar{I}_{xp}$ (red) $\bar{I}_{px}$ (blue) and $\bar{I}_{pp}$ (black) as a function of $\bar{I}_{xx}$ along with fits to $\bar{I}_{AB} = \bar{I}_{xx} +$ constant. The crosses correspond to $\gamma > 1$, while the full circles correspond to $\gamma < 1$. (b) same as (a), but for $\bar{D}_{xp}$ (red) $\bar{D}_{px}$ (blue) and $\bar{D}_{pp}$ (black) as a function of $\bar{D}_{xx}$. Note that the black and the red are on top of each other in (b), which is why the red circles cannot be seen.

size in comparison. From the preceding discussion and Eq. (S.36) they all scale $\propto j^2$. These observations are all confirmed by a full numerical evaluation of $K_{jk}$ and $J_{jk}$. In fact we find that even for relatively small $j$ the matrix elements are well-described by $K_{j,j+1} = K_{j,j-1} \propto -j^2$ and $K_{j,j} \propto j^2$. As implied by Fig. 2 a similar $j^2$ scaling is obtained for $K_{jk}$ involving momentum operators when evaluated numerically. In addition to the finding that $\bar{I}_{xx}$ is closely related to the variance of the work-distribution this analysis also shows that $K_{jk}$ can be effectively described by a tri-diagonal matrix and that $\bar{D}_{xx}$ consequently can be simplified. Indeed, for opening the trap we find that the coefficients are all positive (numerically and for the analytical non-interacting case [5]) which means that the off-diagonal negative values are subtracted from the diagonal giving an overall small value, while for squeezing the trap they alternate between positive and negative values, which adds an overall contribution that also scales as approximately $j^2$, explaining why these values are also proportional to $\Delta W^2$ with a larger proportionality constant than for $\bar{I}_{xx}$.

# CORRELATION FUNCTIONS FOR OTHER COMBINATIONS OF CANONICAL OPERATORS FOR $N = 2$

In the main text we focused on $[\hat{x}_1(t), \hat{x}_1]^2$ as no regularisation of the delta-function is required and as it contains the essential physics of the scrambling. In this section we will argue this claim in more detail by investigating other combinations of canonical operators for the relative Hamiltonian in the two-particle case. We do this utilizing a short-range Gaussian interaction described by $V_{\text{int}}(x) = \kappa \frac{1}{\sigma\sqrt{2\pi}} e^{-x^2/\sigma^2}$. We consider $\kappa = 10$ and $\sigma = 0.04$, which is a situation that has an energy spectrum that corresponds well to $g = 10$ for the delta-interaction. We calculate the time-dependent correlation functions numerically employing the Lagrange-mesh method [6] and find the time-average by considering a time-interval $t \in [1000\pi, 2000\pi]$ which is large enough to get a representative average (see the main text for a further discussion of this). In Fig.2(a) we show $\bar{I}_{AB}$ as a function of $\bar{I}_{xx}$ for a series of quenches where $\gamma \in [1, 20]$ and $\gamma \in [1, 1/20]$, respectively. $\bar{I}_{AB}$ for any combination of operators only differs from $\bar{I}_{xx}$ by a constant as can be seen from the graph where we show fits to $\bar{I}_{AB} = \bar{I}_{xx} +$ constant. The behavior for $\gamma > 1$ and $\gamma < 1$ is also qualitatively similar regardless of the combination of operators. In Fig.2(b) we plot the same, but for $\bar{D}_{AB}$. $\bar{D}_{px}$ and $\bar{D}_{xx}$ (both of which probe the time-dependence of $\hat{x}(t)$) have a linear relationship. For squeezing the trap this results in linear growth of $\bar{D}_{px}$ with $\bar{D}_{xx}$ (blue full circles in the plot), while for $\gamma > 1$ they are both given by a small constant resulting in all points being on top of each other centered at (0,0) in the figure (blue crosses). $\bar{D}_{pp}$ and $\bar{D}_{xp}$ (corresponding to probing the time-dependence of $\hat{p}(t)$) also have a linear relationship. For squeezing the trap $\bar{D}_{pp}$ and $\bar{D}_{xp}$ remain small and constant while $\bar{D}_{xx}$ grows with $\gamma$. For opening the trap on the other hand $\bar{D}_{pp}$ and $\bar{D}_{xp}$ grow with $\gamma$, while $\bar{D}_{xx}$ remains constant. Probing $\hat{x}(t)$ with $\bar{D}_{Ax}$ or $\hat{p}(t)$ with $\bar{D}_{Ap}$ therefore results in the opposite behavior with respect to opening or squeezing the trap, making the overall scrambling different. $\bar{I}_{AB}$, however, always gives a similar contribution to the scrambling, regardless of the combination of canonical operators. Note that regardless of the combination of operators, the infinite time-averaged scrambling is always proportional to the work fluctuations.

---



\end{document}